\documentclass[11pt]{article}

\usepackage{amsmath, amssymb}
\usepackage{graphicx}
\usepackage{booktabs}
\usepackage{hyperref}
\usepackage[a4paper, margin=1in]{geometry}
\usepackage{listings}
\usepackage{xcolor}

\usepackage{rotating}
\usepackage{array}
\usepackage{graphicx}
\usepackage{adjustbox}
\usepackage{float}
\usepackage{tabularx}

\title{On Integrating Resilience and Human Oversight into LLM-Assisted Modeling Workflows for Digital Twins}

\author{
Lekshmi P\\
Indian Institute of Technology Goa\\
Goa, India\\
\texttt{lekshmi20231101@iitgoa.ac.in}
\and
Neha Karanjkar\\
Indian Institute of Technology Goa\\
Goa, India\\
\texttt{nehak@iitgoa.ac.in}
}

\date{}

\begin{document}

\maketitle

\begin{abstract}
LLM-assisted modeling holds the potential to rapidly build executable Digital Twins of complex systems from only coarse descriptions and sensor data. However, resilience to LLM hallucination, human oversight, and real-time model adaptability remain challenging and often mutually conflicting requirements. \textbf{We present three critical design principles} for integrating resilience and oversight into such workflows, derived from insights gained through our work on \textit{\textbf{FactoryFlow}} -- an open-source, LLM-assisted framework for building simulation-based Digital Twins of manufacturing systems.
\textbf{First, orthogonalize structural modeling and parameter fitting.} Structural descriptions (components, interconnections) are LLM-translated from coarse natural language to an\textbf{ intermediate representation (IR) }with human visualization and validation, which is then algorithmically converted to the final model. Parameter inference, in contrast, operates continuously on sensor data streams with expert-tunable controls. \textbf{Second, restrict the model IR} to interconnections of parameterized, pre-validated library components rather than monolithic simulation code, enabling interpretability and error-resilience. \textbf{Third, and most important, is to use a density-preserving IR. }When IR descriptions expand dramatically from compact inputs (turning "100x100 machines as 2D grid" into 10,000 explicit declarations, as XML-netlists do) hallucination errors accumulate proportionally. We present the case for Python as a density-preserving IR: loops express regularity compactly, classes capture hierarchy and composition, and the result remains highly readable while exploiting LLMs' strong code-generation capabilities.
\textbf{A key contribution} is detailed characterization of LLM-induced errors across model descriptions of varying detail and complexity, revealing how IR choice critically impacts error rates. These insights provide actionable guidance for building resilient and transparent LLM-assisted simulation automation workflows.
\end{abstract}

\section{Introduction}\label{sec:introduction}
Digital Twins of manufacturing systems require executable simulation models that can adapt continuously to real-time sensor data and evolving production environments. Traditional modeling and simulation (M\&S) workflows (consisting of sequential stages of measurement, input modeling, model specification, conversion to executable form, verification, and validation) are too slow and rigid for critical applications where production systems evolve rapidly. For manufacturing system Digital Twins (DTs), where timely decision-support and optimization are essential, this traditional cycle is increasingly obsolete.

Two key driving forces make automated model generation not only attractive, but imperative. First, the proliferation of IoT sensors and Industry 4.0 initiatives has created an abundance of real-time operational data from manufacturing systems. Second, advances in cloud and edge computing, low-latency networks, and machine learning have made it feasible to leverage AI for automating the entire modeling lifecycle: initial model generation, continuous fitting and tuning to sensor data, ongoing monitoring and adaptation, and model-based decision support including what-if analysis, optimization, and control~\cite{qiu2025review, kalsoom2020advances}.

\textbf{Automated Model Generation (AMG)} emerged as an active research area in the early 2010s, with approaches focusing on process mining techniques to infer system behavior from event logs and flows/structure from sensor data, producing models such as Petri nets and Markov chains~\cite{friederich2022process, friederich2022framework}. Purely data-driven methods were developed to infer component parameters and system structure directly from operational data~\cite{tan2023automatic}. More recently, the dramatic advances in Large Language Models (LLMs) have opened new possibilities: using LLMs as modeling assistants or even as model representations themselves. The ability of LLMs to translate natural language descriptions into formal representations has made LLM-assisted automation increasingly attractive for simulation modeling workflows~\cite{alexiadis2024text,kumar2025performance,LLM_SD_WSC_2025}.

However, practical applicability and trust in LLM-assisted automation remain contingent on overcoming inherent issues: hallucination (generation of plausible but incorrect outputs), computational cost and API stability of commercial models, lack of transparency, and alignment challenges~\cite{huang2024survey, alansari2025hallucination}. \textit{While these fundamental limitations in LLMs may be addressed in future, practical applications today require workflows that systematically work around these issues through deliberate design principles.} This is our approach.

\textbf{Problem Definition and Scope:} In this paper, we narrow our scope to a specific but important class of 
systems: discrete-event simulation (DES) models of manufacturing systems 
where parameters (such as machine task delays, energy profiles, state transition rates, failure rates) evolve continuously but structural changes occur sporadically. For example, changes to equipment interconnections, active machines, job sequencing and routing.  This characterization is typical of manufacturing systems 
such as assembly lines, packaging lines, semiconductor fabrication facilities, and consumer goods production plants~\cite{may2024automated}. While parameters evolve continuously with operational conditions, structural changes are infrequent and typically planned (equipment upgrades, line reconfigurations, capacity expansions). This scoping creates valuable opportunities. First, \textit{orthogonalization becomes possible}: structural modeling and parameter inference require fundamentally different automation strategies and can be decoupled. Structural descriptions benefit from expert knowledge about system topology and can be updated sporadically (with expert-in-loop, either when a change is detected automatically or planned to occur deliberately); while parameter fitting must operate continuously on streaming sensor data. Second, \textit{expert-in-the-loop workflows become practical}: for parameter inference, experts can configure data filters, time windows, distribution families, and mappings between sensor streams and model parameters through graphical interfaces, with automated fitting running continuously under these constraints. For structural modeling, factory operators and production engineers (who may possess deep knowledge of their systems but may lack M\&S expertise) can describe system structure in natural language, with LLM-assisted translation producing formal models subject to human (visual) and automated rule-based or test-based validation. This democratization of modeling enables domain experts to contribute their knowledge directly, avoiding pitfalls of purely data-driven approaches to model building where data is scarce, stale, or contains outliers that require human/operator interpretation to handle appropriately. Despite these opportunities, critical challenges remain for making LLM-assisted automation trustworthy in practice: 

    \textbf{First is the challenge of resilience} to errors such as hallucination. When LLMs generate monolithic simulation code directly (for example, ~\cite{LLM_SD_WSC_2025,kumar2025performance,erika2024}), hallucination can introduce subtle errors, some of which may manifest only during execution; in the best case causing simulation crashes, and in the worst case producing plausible but incorrect predictions that lead to flawed decisions. 
    %
    \textbf{Second is the challenge of systematic human oversight}. True expert-in-the-loop workflows require more than occasional human validation; they demand systematic integration of domain expertise throughout the entire process. The workflow must enable experts to contribute their knowledge about the system while remaining accessible to users without specialized M\&S training.
    These two challenges are interconnected: oversight mechanisms enable resilience by allowing human validation to catch errors, while resilience to errors enables effective oversight. Both require deliberate architectural choices in how automation workflows are designed.

\subsection{Main Contributions}
This paper presents three critical design principles for integrating resilience
and human oversight into LLM-assisted modeling workflows. These
principles emerged from the iterative development and refinement
of \textit{ FactoryFlow}, our open-source framework for building simulation-based Digital Twins of manufacturing systems. FactoryFlow combines LLM-assisted structural modeling, real-time parameter inference from sensor data, and systematic expert validation to produce executable discrete-event simulation models. FactoryFlow is publicly released as open source at \textcolor{blue}{\url{https://github.com/InferaFactorySim/FactoryFlow}}~\cite{FactoryFlowGitHub}.

\textbf{Previous Work:} An initial prototype presented at WinterSim 2025~\cite{WSC2025_FactoryFlow} introduced the conceptual framework and presented \textbf{FactorySimPy} (a validated component simulation library) as the main contribution, along with a proof-of-concept LLM-translation implemented using a \textbf{netlist-like IR} (where each component instance was explicitly listed out as a dictionary along with its interconnections). 
Subsequently, through analysis and iterative improvement of that implementation, we observed that hallucination errors occurred more frequently when compact natural language descriptions expanded into large netlist 
representations. This led us to explore alternative IRs, including XML and JSON formats for netlists, and ultimately Python for structural description with its native support for loops and classes to represent regular structures, composition and hierarchy. Several examples also revealed various error types beyond hallucination in LLM-generated descriptions. We found that \textit{a systematic error characterization for LLM-assisted simulation model generation is largely absent in published literature}, motivating a key focus of this work.
Building on this, the current paper makes the following contributions:

\begin{enumerate}
\item \textbf{Three design principles for resilient and transparent LLM-assisted 
workflows} (particularly suited to the manufacturing DTs context). We present, justify, and demonstrate three critical principles: 
 (1) orthogonalization of structural modeling and parameter fitting with systematic human-in-the-loop integration, (2) component-based composition rather than monolithic code generation, and (3) density-preserving intermediate representation design. We justify each principle giving examples and describe its implementation in FactoryFlow. For principle 3, we empirically demonstrate that a \textbf{density-preserving IR} (specifically Python) reduces hallucination errors relative to XML-netlists by enabling compact, readable representations through loops and classes, while leveraging strong code-generation capabilities in modern LLMs.

\item \textbf{Detailed error characterization and actionable insights.} Through 
systematic analysis of LLM-induced errors across models of varying size and topological 
complexity, we characterize error types, their frequencies, and how they depend on 
model characteristics and IR choice. From this empirical analysis, we derive actionable 
insights for building trustworthy LLM-assisted simulation automation workflows, with 
principles generalizable beyond manufacturing Digital Twins.
\end{enumerate}

The remainder of this paper reviews related work to place our contributions in context (Section~\ref{sec:related}), presents a summary of the FactoryFlow architecture (Section~\ref{sec:factoryflow}), and the three design principles using examples and the case for Python as an IR (Section~\ref{sec:principles}). Section~\ref{sec:results} describes our experimental methodology, presents error characterization results, and discusses key insights and broader implications. Section~\ref{sec:conclusion} concludes with a discussion of unresolved challenges, limitations in current implementation, takeaways and future directions.

\section{Related Work} \label{sec:related}
\textbf{Automated Simulation Model Generation (AMG \slash ASMG)} emerged well before large language models,
rooted in process mining, knowledge-based methods, and data-driven parameter inference~\cite{lugaresi2021automated, lugaresi2023automated, CIMINO_ASMG_LR_2025, 
C2AC_TOMACS_2024}. Process mining techniques infer system behavior from event logs, 
producing Petri nets and enabling automated digital twin generation~\cite{lugaresi2021automated, 
Lugaresi2024process}. Data-driven approaches extract parameters from operational sensor 
data~\cite{tan2023automatic}. These foundational methods established the importance 
of separating structural discovery from parameter tuning, a principle that remains 
relevant as LLMs introduce new automation capabilities.

\textbf{LLMs for simulation model generation.}
Large language models have opened new avenues through natural language interfaces 
supporting the entire simulation lifecycle~\cite{GPTbasedSimulation_WSC2023}. Direct 
natural language to code approaches translate textual descriptions into executable 
models~\cite{kumar2025performance, JacksonIvanov2024}, including generation for 
proprietary platforms (FlexSim~\cite{ChatGPT2FLEXSIM_2026}, DEVS~\cite{DEVScopilot_WSC2024}), 
domain-specific languages, and System Dynamics specifications~\cite{LLM_SD_WSC_2025}. Multi-agent systems employ LLMs to configure agent behaviors~\cite{LLM_MultiAgent_hattori_WSC2024}, spanning agent-based models~\cite{LLM_ABM_WSC_2024, 
NetLogo_chatGPT_CHI_ABM_2024} and manufacturing planning~\cite{NLP2PRODPLAN_2025}. While 
democratizing modeling for non-experts, these approaches often require human-in-the-loop 
debugging to correct logic errors and hallucinations~\cite{kumar2025performance, 
LLM_SD_WSC_2025, huang2024survey, alansari2025hallucination}.

\textbf{Intermediate representations and structured generation.}
To improve reliability, structured approaches employ intermediate representations 
bridging natural language and executable artifacts. Template-guided methods constrain 
LLM outputs to predefined schemas: conversational knowledge extraction populates 
templates that instantiate models directly~\cite{NLP2PRODPLAN_2025}, 
while formal modeling languages with scalable templates ensure structural and semantic 
correctness~\cite{LLM2X_SIMPAT2026}. Schema-based representations including CMSD, 
XML, and JSON provide standardized formats for manufacturing simulation~\cite{may2024automated, 
AI4factory_2025}, though their instance-oriented nature limits scalability for large, 
regular structures. Domain-specific languages with grammar constrained generation 
reduce syntax errors through few-shot prompting and constrained decoding~\cite{LLM_AMG_DSL_2025_TOMACS}. 
Multi-agent systems have translated specifications into synthesizable hardware design 
code~\cite{yu2025spec2rtl}. These IR-based approaches trade flexibility for correctness 
guarantees, though most still require domain expertise to validate intermediate outputs.

\textbf{Verification, validation, and error characterization.}
Reliability depends on systematic verification mechanisms. Grammar constrained template 
filling significantly reduces syntax errors~\cite{LLM2X_SIMPAT2026}, while FSM-based 
checks ensure logical feasibility of manufacturing workflows~\cite{kumar2025performance, 
LLM4PROCESSPLAN_2025}. Formal error characterization through metrics such as Degree 
of Error and Model Consistency quantifies practical impact~\cite{LLM2X_SIMPAT2026}. 
Agentic frameworks demonstrate self-validation where LLM agents generate models, 
conduct in-silico experiments, and perform self-evaluation~\cite{moeltner2026}. A recent taxonomy of LLM code-generation errors~\cite{error_taxonomy_ICSE_2025} categorizes failures into semantic and syntactic classes and reports that most stem from reasoning about task requirements rather than from syntactic mistakes. Our eight-category taxonomy is consistent with this finding in the DES modeling setting: syntactic errors (T7) were absent across all 35 benchmark models, while semantic errors dominated, particularly structural hallucinations (T3, T4) and hierarchy mismatches (T6).

Our work addresses gaps in existing approaches: (1) orthogonalization of structural 
modeling from continuous parameter fitting in operational systems, (2) systematic 
characterization of error types in LLM-generated DES models, and (3) design principles 
for intermediate representations that preserve information density while enabling 
human oversight, all with alignment to the application area of manufacturing DTs.

\section{FactoryFlow Architecture} \label{sec:factoryflow}
\begin{figure*}[btp]
  \begin{center}

  \includegraphics[width=0.9\textwidth]{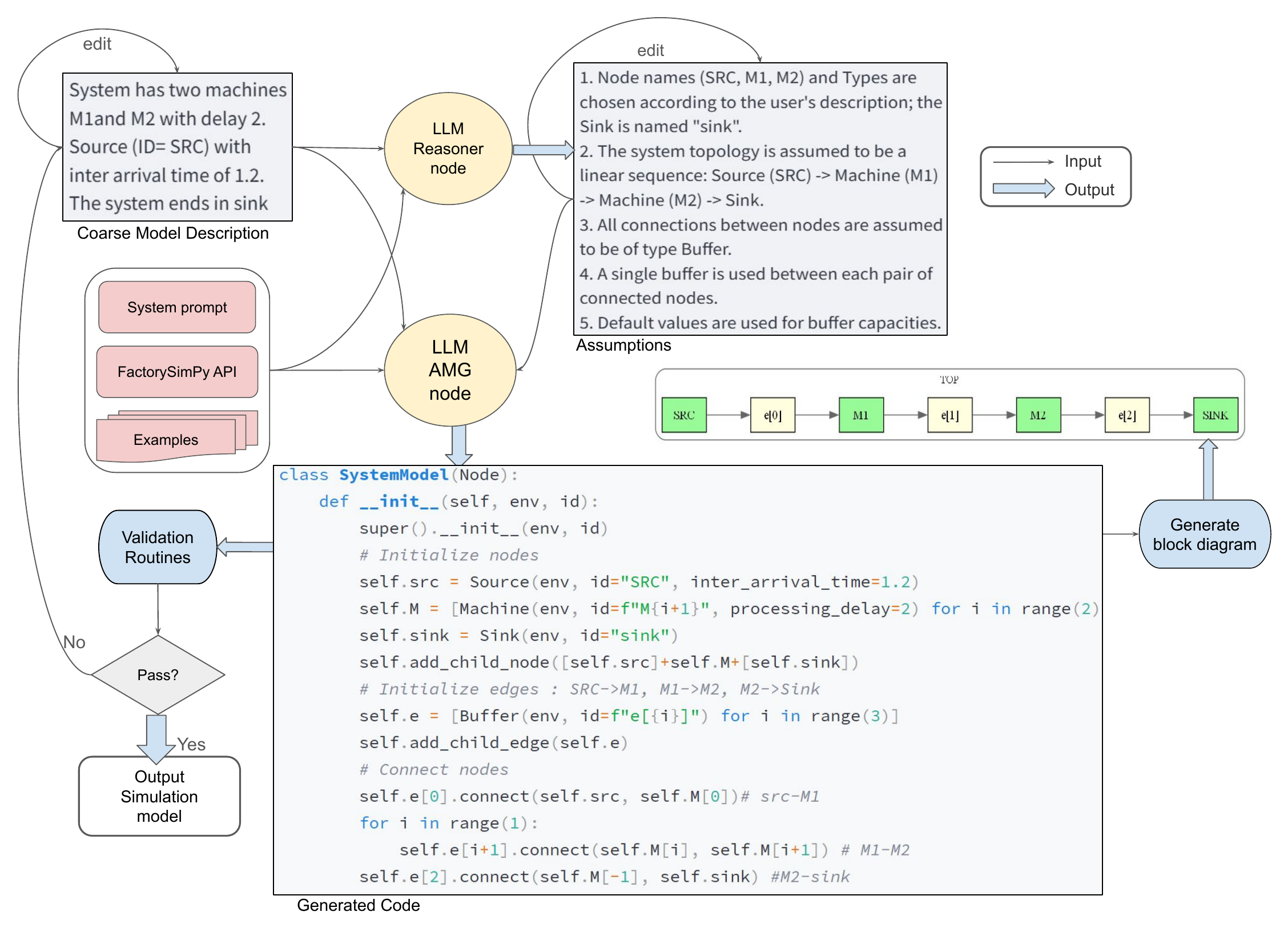}
  \caption{LangGraph-based architecture for LLM-assisted structural model generation in FactoryFlow}
  \label{fig:architecture}
  \end{center}
\end{figure*}

FactoryFlow is an open-source framework for generating executable discrete-event 
simulation models of manufacturing Digital Twins from coarse natural language 
descriptions and sensor data. The framework comprises three main components: 
\emph{DataFITR} for real-time parameter inference and distribution fitting from 
sensor streams, \emph{FactorySimPy} as a validated library of configurable 
manufacturing system components, and an \emph{LLM-based structural model generator} 
that translates natural language descriptions into component interconnections~\cite{FactorySimPyGitHub,FactorySimPyDocumentation}.

The architecture and initial prototype implementation were described 
in~\cite{WSC2025_FactoryFlow}. While that work focused on developing the core 
FactorySimPy library and a prototype LLM flow using netlist-like representations, 
the current paper presents a refined LLM-based structural modeling architecture.



The LLM-based structural modeling component uses a LangGraph architecture with two 
primary nodes (Figure~\ref{fig:architecture}). The \emph{reasoning node} interprets 
the system description, identifies missing information, and proposes explicit assumptions 
(component identifiers, buffer counts, routing policies). The \emph{code-generation node} 
receives the description, assumptions, and supporting materials (FactorySimPy API 
specification, example models, system prompt) to generate executable Python code by 
instantiating FactorySimPy components and their interconnections. 
We use Gemini 2.5 Pro with few-shot prompting 
that demonstrates component instantiation, interconnection patterns, parameter passing, 
and naming conventions. FactoryFlow maintains structured state containing the user description and inferred 
assumptions, updated incrementally across LangGraph iterations. 

\textbf{Generated Model and Validation Support:} The generated model is in the form of Python code that imports FactorySimPy,
instantiates components from the library (such as machines, conveyor belts and buffers) and initializes
their parameters and interconnections (similar to model descriptions built using commercial tools such as Arena or FlexSim). It is returned to the user along with an automatically derived block diagram for visual inspection. The user has multiple options for refinement: directly modify the generated Python code, 
edit the original natural language description, adjust the inferred assumptions, 
or any combination thereof. Modified inputs can be resubmitted to the LangGraph 
pipeline, enabling iterative refinement until the model satisfies domain requirements.

Once the user accepts the generated model, automated validation routines execute
to detect structural errors including isolated components, missing connections,
unspecified mandatory parameters, and violations of component interconnection rules
defined in the FactorySimPy library. These validation checks provide an additional
layer of error detection beyond human inspection, ensuring that only well-formed
models proceed to execution.

\textbf{Target users and expected interaction:} FactoryFlow targets factory operators, production engineers, and simulation experts. Non-M\&S users are not expected to read or write the Python IR. Their typical interaction is through the natural-language description and the automatically generated block diagram, with LLM assistance handling IR synthesis. Simulation experts may additionally inspect or edit the IR directly. Extending the behavior of components, as opposed to their composition, requires modifying the FactorySimPy library and is an M\&S-expert workflow.

Section~\ref{sec:principles} describes the three design principles and how they are incorporated
into this architecture.


\section{Design Principles and IR Choice} \label{sec:principles}

We present three critical design principles for integrating resilience and human 
oversight into LLM-assisted modeling workflows. These principles are particularly 
suited to manufacturing Digital Twins where system structure changes sporadically 
(equipment layout, interconnections, routing) while parameters evolve continuously 
(task delays, energy profiles, machine states).

\subsubsection*{\textbf{Principle 1: Orthogonalize Structural Modeling and Parameter Fitting}}

Structural modeling and parameter inference serve fundamentally different purposes 
and benefit from independent automation strategies. In FactoryFlow, this 
orthogonalization is realized through two separate subsystems.

\textbf{Parameter fitting (DataFITR):} Parameter inference operates continuously 
on real-time sensor streams using distribution fitting techniques that identify 
appropriate distribution families and estimate parameters. LLMs are not required; 
classical statistical methods and ML algorithms suited to time-series analysis 
suffice. The process is expert-tunable through GUI-based controls where users 
configure time windows, select candidate distribution families, and specify 
mappings between sensor streams and model parameters. Random variates generated 
by DataFITR are automatically matched to component parameters by instance name, 
with manual configuration available when needed. Once configured, algorithms 
perform continuous fitting, updating model parameters automatically.

\textbf{Structural modeling (LLM-assisted):} System structure (components, 
interconnections, routing policies) is translated from coarse natural language 
using the LLM-based flow described in Section~\ref{sec:factoryflow}. Validation 
occurs through automatically generated block diagrams and built-in routines in 
FactorySimPy that check for isolated components, missing connections, and 
unspecified parameters. When structural changes occur, expert input is solicited 
to modify the description and regenerate the model. Additional validation can 
compare known performance metrics such as cycle time between the real system and 
simulation.


Elevating this separation to a design principle rests on a contrast with existing AMG approaches and on an asymmetry inherent to manufacturing systems. Several data-driven AMG methods infer structure and parameters jointly from a single source: process mining derives both topology and timing/routing parameters from event logs, and Petri-net or FSM discovery methods learn transitions and rates together. Such combined approaches fit settings where topology itself must be discovered from data. In the manufacturing DT setting, structural changes are sporadic and planned (line reconfigurations, equipment upgrades) and are most reliably captured from expert natural-language descriptions, while parameter drift is continuous (tool wear, throughput variability, failure-rate shifts) and is most reliably captured from live sensor streams. Binding the two into a single automation step forces a compromise on whichever axis is less well matched to the chosen data source. Orthogonalization lets each use the strategy best suited to its timescale and data source: LLM-assisted synthesis with expert validation for structural updates, statistical fitting for parameter updates. A secondary benefit is that the structural generator and DataFITR can be validated, tuned, and evolved independently.

\subsubsection*{\textbf{Principle 2: Component-Based Composition}}

Rather than having LLMs generate monolithic simulation code from scratch~\cite{LLM_SD_WSC_2025, 
kumar2025performance, erika2024}, FactoryFlow constrains model generation to 
instantiating and interconnecting validated components from the FactorySimPy library. 
FactorySimPy provides pre-validated, well-documented component classes for elements 
commonly found in manufacturing systems: machines with various processing policies, 
buffers with configurable capacities and queueing disciplines, conveyors, material 
handling fleets, splitters, mergers, and sources/sinks. While minimal in scope, 
the library is designed for extensibility.

This architectural constraint yields several benefits. Errors can only occur in 
component instantiation and interconnection, not in underlying simulation mechanics, 
as FactorySimPy handles all low-level concerns. Automated validation routines check for common structural 
errors such as isolated components, missing connections, duplicate instance names, 
and violations of interconnection rules. More subtle bugs such as race conditions 
cannot result from model construction, as component semantics enforce proper 
synchronization. Unlike subtle deadlocks that can be caused in monolithic simulation code, 
if a purely structural description produces deadlock, this often indicates a real 
deadlock possibility in the actual system. Finally, models composed of familiar, 
named components (\texttt{Machine\_A}, \texttt{Buffer\_B}) are far more interpretable 
to domain experts than monolithic code with complex event scheduling.

\textbf{Tradeoff:} Component-based modeling is more rigid. Behaviors not supported
by existing components require library extension. This approach suits domains where
a canonical set of building blocks is known and relatively stable, as is typical
in manufacturing systems.
This is reflected in our error characterization (Section~\ref{sec:results}): no Python syntax errors (T7) were observed across the 35 benchmark models, and residual errors were confined to component naming, parameter assignment, structural hallucinations, and framework-level constraints.

\subsubsection*{\textbf{Principle 3: Density-Preserving Intermediate Representations}}

This principle addresses how system structure is represented in the intermediate 
form generated by the LLM. Traditional structural descriptions rely on enumerative 
formats where each component instance and connection is explicitly listed (netlists 
in circuits, UML in software, XML/JSON in data formats). While adequate for small 
systems, these formats exhibit critical weaknesses when produced by LLMs for large 
or regular structures.

\textbf{The Problem: Enumerative Expansion and Error Accumulation:} Consider the 
description: ``Create a 100$\times$100 grid of machines, each connected 
to its four nearest neighbors.'' This compact input describes 10,000 machines and 
approximately 40,000 connections. In an enumerative netlist-based IR, the LLM must 
explicitly generate entries for all 10,000 machines and 40,000 connections, each 
requiring correct identifiers, parameters, and connection endpoints.
We observed that hallucination errors accumulate proportionally with this expansion. 
Common errors in netlist-based IR included naming inconsistencies (\texttt{machine\_50\_50} 
versus \texttt{Machine\_50\_50}), off-by-one errors in indexing, missing boundary 
connections, hallucinated parameter values, and duplicate declarations. Furthermore, 
these massive netlists (thousands of lines) become unreadable to human experts, 
defeating the goal of enabling oversight.

\textbf{The Solution: Density-Preserving IRs:} Intermediate representations 
should maintain proportional complexity to natural language inputs by supporting 
programmatic constructs for modularity (functions and classes), hierarchy 
(composition mechanisms), and regular structures (loops, comprehensions, generators).
An IR with these features allows ``100$\times$100 mesh'' to remain compact. For 
the mesh example, a density-preserving Python IR might be:
\begin{verbatim}
machines = [[Machine(f"m_{i}_{j}") for j in range(100)] 
    for i in range(100)]
for i in range(100):
    for j in range(100):
        if j < 99: machines[i][j].connect(machines[i][j+1])
        if i < 99: machines[i][j].connect(machines[i+1][j])
\end{verbatim}
This representation is approximately 5 lines regardless of grid size, versus 
50,000+ lines for enumerative netlists. The loop-based structure mirrors regularity 
in the natural language description, reducing hallucination opportunities while 
remaining readable.
This principle is not unique to Python. Hardware description languages (VHDL, 
Verilog) provide \texttt{generate} statements and parameterized modules precisely 
to enable compact specification of regular structures, reducing human error. This 
becomes even more critical when the generator is an LLM prone to inconsistencies.

Python is well-suited as a density-preserving IR because it provides native support 
for loops and comprehensions (enabling compact expression of regular topologies), 
a class system supporting hierarchical organization, high readability when constrained 
to component instantiation patterns, strong LLM code-generation capabilities (Python 
is heavily represented in training corpora), and natural alignment with FactorySimPy 
(which is Python-based).
Our empirical evaluation (Section~\ref{sec:results}) demonstrates that the hallucination errors
accumulate in proportion to the size of the LLM's output, which in turn depends on the choice of the IR.
We observed that transition from netlist to Python IR substantially reduced hallucination error counts across models
of varying complexity.
Beyond the benchmark set, we have used FactoryFlow to generate hierarchical manufacturing lines with regular structures containing more than 10000 components. The generated IR remains a few tens of lines in these cases, and the natural-language descriptions and FactoryFlow-generated models are available in the GitHub repository.

\section{Error Characterization Results and Insights} \label{sec:results}

This section presents empirical analysis of errors in LLM-generated simulation 
models across varying system sizes and description granularities. A comprehensive 
dataset documenting this characterization study including the benchmark model set, 
natural language descriptions, LLM-generated outputs, identified errors, and 
analysis scripts is \textbf{publicly available} in the FactoryFlow GitHub repository~\cite{FactoryFlowGitHub}
at \href{https://github.com/InferaFactorySim/FactoryFlow/tree/main/docs/error-characterization-results}{this link}.

\subsection{Methodology} 
We constructed a benchmark set of 35 manufacturing system models (S1-S35) with 
varying sizes and topological complexity. System size ranges from simple serial 
configurations with fewer than 10 components to large hierarchical systems with 
100+ components. The set includes:

\begin{itemize}
\item Simple serial systems (S1-S5)
\item Parallel systems and feedback loops (S6-S11)
\item Multi-edge systems with routing policies (S8, S23, S25, S27)
\item Hierarchical and nested subsystems (S12, S18, S26, S29, S30)
\item Irregular or heterogeneous interconnections (S21, S22)
\item Very large systems with regular structures (S24, S31-S35)
\end{itemize}

For each system, we prepared three artifacts: (1) a \emph{ground-truth} implementation 
written in Python using FactorySimPy, representing the intended system structure 
with correct component names, connectivity, and parameter values; (2) a \emph{coarse 
natural language description} providing high-level system overview with minimal or 
no specification of component identifiers, parameter values, or naming conventions; 
and (3) a \emph{detailed natural language description} fully specifying system 
structure including component identifiers, parameters, connectivity, and prescribed 
naming patterns.

Each description (coarse and detailed) was processed through FactoryFlow's LLM-based 
pipeline, generating assumptions and intermediate representation (IR) code. The 
generated FactorySimPy models were compared against ground-truth implementations 
at the component, connection, and parameter levels. Errors were identified and 
classified according to the taxonomy described below. Model comparison was carried 
out at the structural and semantic level rather than through exact code matching, 
such that functionally equivalent implementations are treated as correct.
%

\subsection{Error Taxonomy}

Observed errors are classified into eight types grouped under three broad categories: 
lexical, structural, and formal errors. Table~\ref{tab:error_taxonomy} summarizes 
the error types with definitions and representative examples.

\textbf{Lexical errors (T1-T2)} stem from incorrect naming or parameter assignment. 
Naming errors (T1) include inconsistent identifier formats (\texttt{machine\_50\_50} 
versus \texttt{Machine\_50\_50}), off-by-one indexing mistakes where 0-based indexing 
from few-shot examples conflicts with 1-based references in natural language 
descriptions, and misapplied naming conventions. These errors are particularly 
prevalent in grid-based systems and when naming rules depend on spatial attributes 
or procedural generation. Parameter errors (T2) involve incorrect parameter values, 
misapplication of default values when descriptions are ambiguous, or incorrect 
resolution of conflicting parameter specifications.

\textbf{Structural errors (T3-T6)} arise from hallucinations or incorrect structural 
reasoning by the LLM, resulting in divergence between the generated model and 
intended system semantics. Node hallucinations (T3) include addition or omission 
of components such as machines, buffers, sources, or sinks. Edge hallucinations 
(T4) involve incorrect connections between components. Parameter hallucinations 
(T5) occur when the LLM generates parameter specifications not present in the 
description or infers incorrect parameter structures (e.g., routing policies as 
explicit lists rather than supported policy types). Hierarchy mismatches (T6) 
include flattening of nested subsystems, misplacing components across hierarchical 
scopes, or collapsing intended modular structures into global topologies.

\textbf{Formal errors (T7-T8)} correspond to violations of Python syntax (T7) or 
FactorySimPy-specific constraints (T8). Python syntax errors include malformed 
expressions, undefined variables, or improper indentation. FactorySimPy constraint 
violations include invalid edge cardinality (e.g., attempting one-to-many connections 
where one-to-one is required), incompatible port type connections, invalid routing 
policy specifications, or structural patterns that leave components disconnected.

\begin{table*}[htbp]
\caption{Error taxonomy with definitions and examples.}
\label{tab:error_taxonomy}
\small
\begin{tabularx}{\textwidth}{@{}llX@{}}
\toprule
\textbf{ID} & \textbf{Category} & \textbf{Description \& Example} \\
\midrule
T1 & Naming errors & Inconsistent identifiers, off-by-one indexing, or violated 
naming conventions. \emph{Example:} Grid systems use 0-based indexing when description 
specifies 1-based (S21, S24); buffers renamed to generic identifiers from few-shot 
examples (\texttt{edge[0]}, \texttt{src\_edge[0]}); naming errors increase with 
description detail rather than decrease. \\
\addlinespace
T2 & Parameter errors & Incorrect parameter values, misapplied defaults, or ambiguous 
conflict resolution. \emph{Example:} S13 prioritized stage-specific values over generic 
statements; routing policies defaulted to \texttt{ROUND\_ROBIN} when unspecified 
(S30, S31-S35). \\
\addlinespace
\midrule
T3 & Node hallucination & Addition or omission of nodes (machines, buffers, sources, sinks). 
\emph{Example:} S3 hallucinated an extra machine to satisfy perceived edge constraints; 
S35 hallucinated 10 nodes not in description. \\
\addlinespace
T4 & Edge hallucination & Addition or omission of edges (connections between components). 
\emph{Example:} S18 inferred two feedback connections instead of one based on ambiguous 
phrasing; S35 hallucinated 7 edges. \\
\addlinespace
T5 & Parameter hallucination & Generation of parameter specifications not present in 
description or incorrect parameter structures. \emph{Example:} S23, S25 hallucinated 
routing behavior as explicit lists instead of supported policy types; S35 hallucinated 
many parameter values not specified. \\
\addlinespace
T6 & Hierarchy mismatch & Incorrect hierarchical structure, flattening of nested 
subsystems, or misplaced component scope. \emph{Example:} S13, S15, S18, S24, S31 
collapsed explicitly described subsystems into flat topologies; S26 reinterpreted 
hierarchical system as single serial flow; S24 assumed single global source/sink 
instead of per-row hierarchy. \\
\addlinespace
\midrule
T7 & Python syntax & Malformed expressions, undefined variables, indentation errors. 
\emph{Note:} No Python syntax errors observed across all experiments. \\
\addlinespace
T8 & FactorySimPy violations & Invalid edge cardinality, incompatible connections, 
or unsupported specifications. \emph{Example:} S13, S15 attempted one-to-many edges 
violating one-to-one constraint (single buffer connecting source to multiple machines); 
S30 left machine without outgoing edge. \\
\bottomrule
\end{tabularx}
\end{table*}

\subsection{Quantitative Results}

We analyze the LLM-generated models to address \textbf{ four key questions:}
(1) How do error counts correlate with model size and complexity? (2) Which error types are most frequent, and does this vary between coarse and detailed descriptions? (3) What is the impact of description granularity on error profiles? (4) Are certain model characteristics or description styles more error-prone?

\begin{figure}
  \centering
  \includegraphics[width=0.75\textwidth]{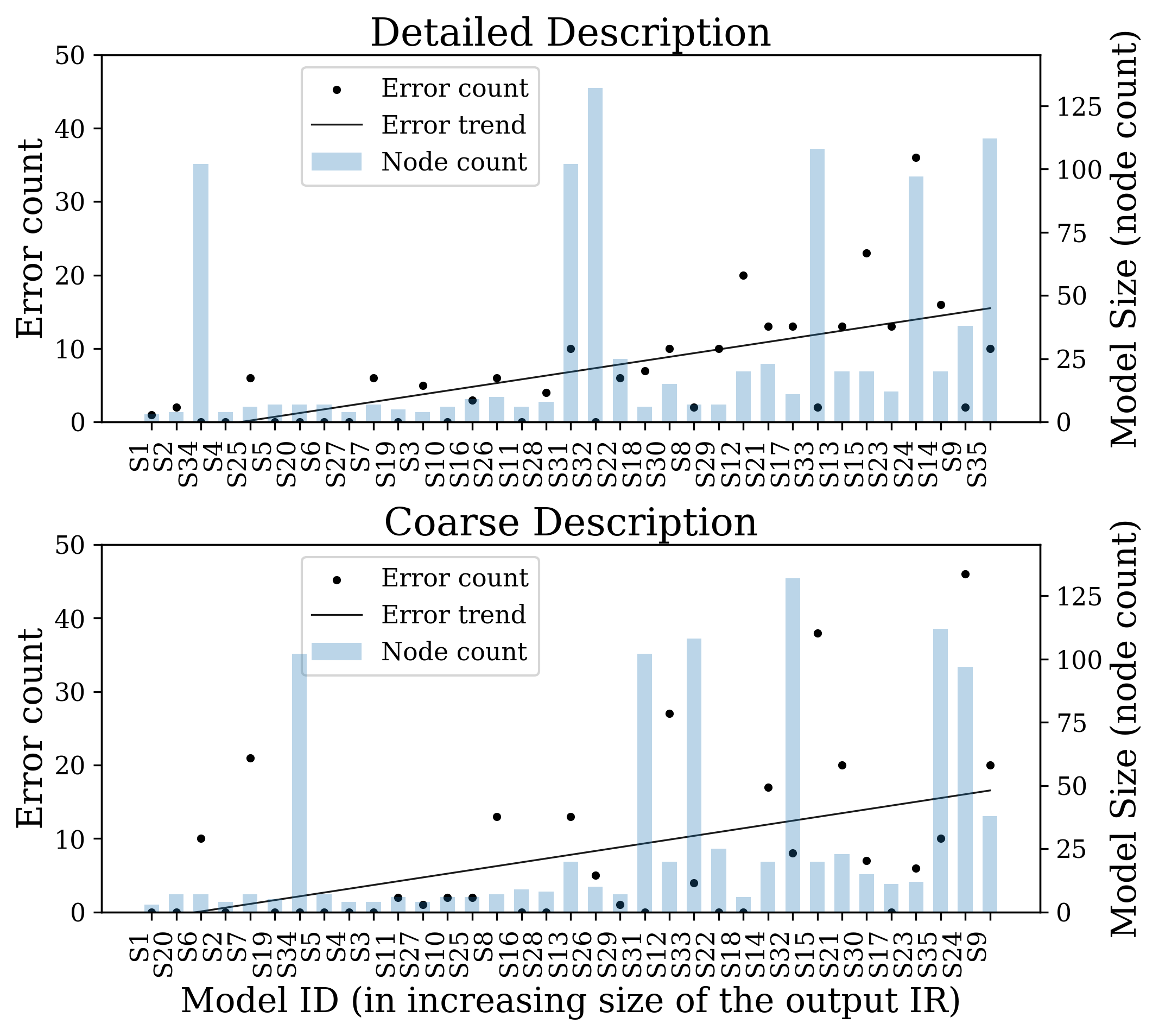}
  \caption{Error counts across models ordered by complexity (IR size).}
  \label{fig:errorVmodel}
\end{figure}

\begin{figure}
  \centering
  \includegraphics[width=0.8\textwidth]{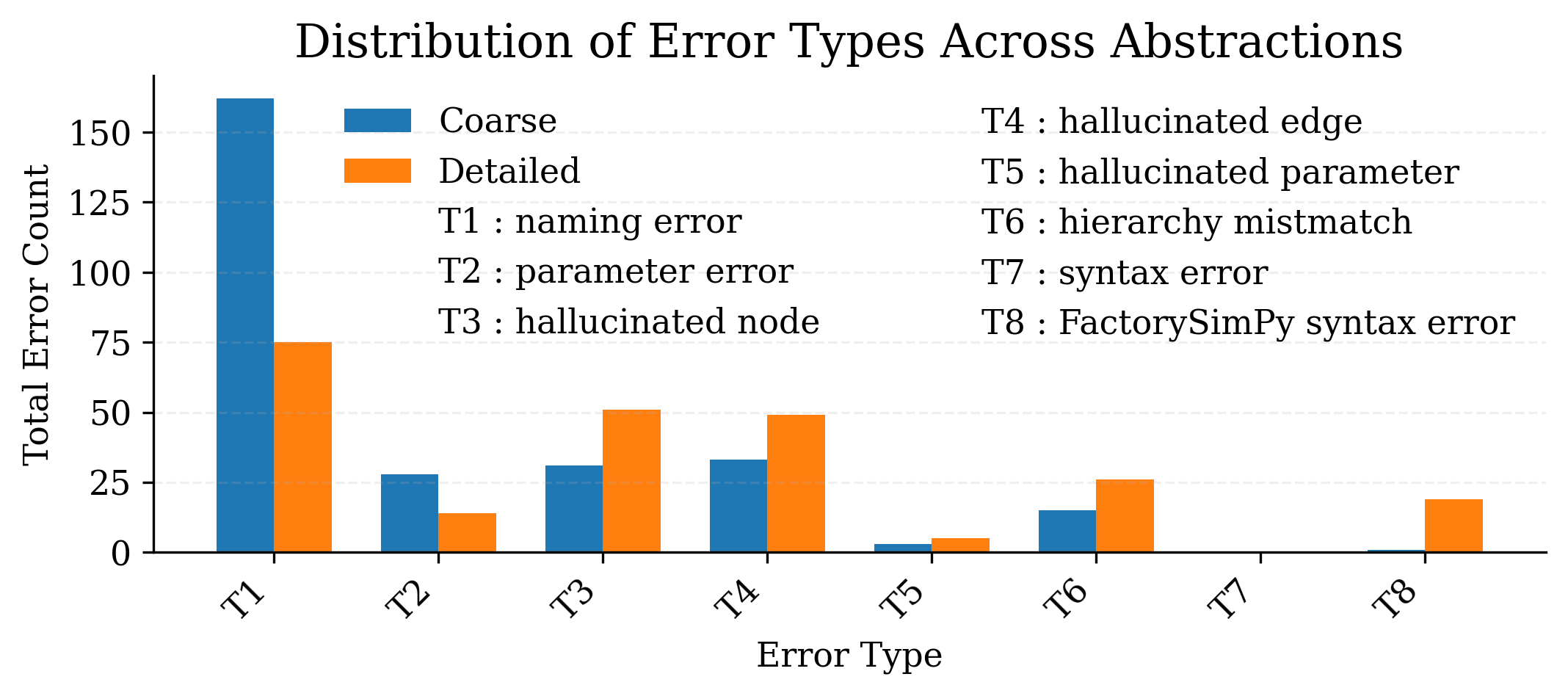}
  \caption{Aggregate error type frequency across all models.}
  \label{fig:error_type_frequency}
\end{figure}

\begin{figure*}[htbp]
  \centering
  \includegraphics[width=0.99\textwidth]{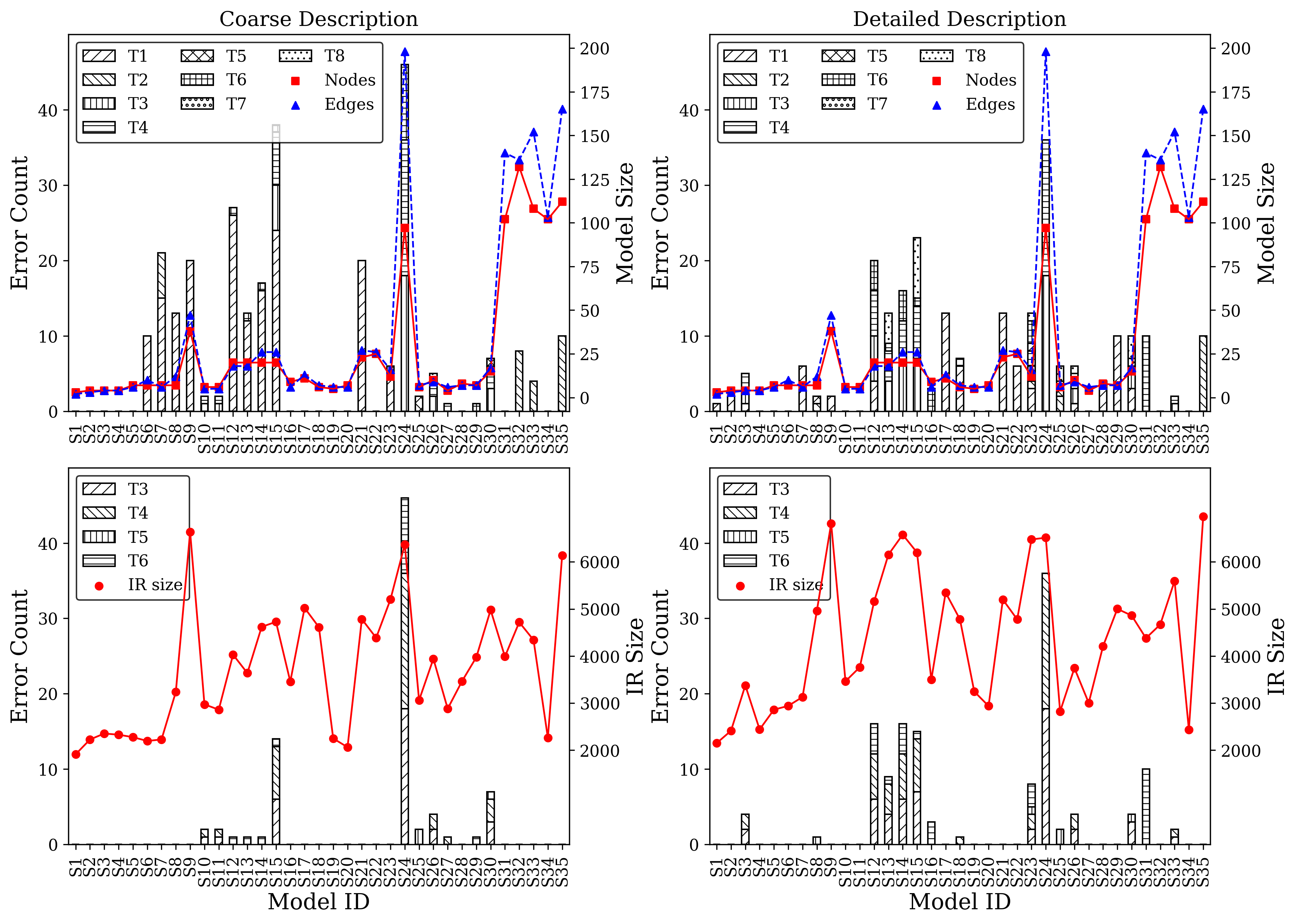}
  \caption{Error composition across individual models for both coarse and detailed descriptions.}
  \label{fig:error_comp}
\end{figure*}

\textbf{Error growth with model size and complexity.}
Figure~\ref{fig:errorVmodel} orders models by increasing complexity, approximated 
by the character count of the generated IR. For each model, bar height represents 
model size (number of component instances), while points denote error counts. A general 
increasing trend shows correlation between model size, IR size, and error counts. 
Smaller models (S1-S5) are typically generated with few or no errors, while larger 
and more structurally complex models exhibit higher error counts. This trend holds 
for both coarse and detailed descriptions. Notably, models S31-S35, despite their 
large size (100+ components), show comparatively lower error rates due to their 
regular, repeated structures. This suggests that structural regularity mitigates 
error accumulation even at scale when using density-preserving IR.

\textbf{Error type distribution and composition.}
Figure~\ref{fig:error_type_frequency} presents aggregate error frequency across all 
models, comparing coarse and detailed descriptions. Naming errors (T1) are most 
frequent under coarse descriptions. It is important to note that coarse descriptions 
do not specify component names, so any generated name is valid in principle; we count 
mismatches with intended ground-truth names to maintain consistent comparison. Detailed 
descriptions reduce T1 but increase T3 (node hallucination), T4 (edge hallucination), 
T6 (hierarchy mismatch), and T8 (FactorySimPy violations).

Figure~\ref{fig:error_comp} shows error composition across individual models using 
stacked bars. For coarse descriptions, naming errors (T1) dominate in medium-to-large 
models (S7-S15, S21, S24), reflecting identifier inference ambiguity. Node and edge 
hallucinations (T3, T4) appear frequently in structurally irregular systems (S30, S24). 
Hierarchy errors (T6) emerge in systems with repeated or nested structures but remain 
less frequent overall. For detailed descriptions, naming errors are reduced in smaller 
systems, but hierarchy errors (T6) become more pronounced in models with repeated 
subsystems or cross-connections (S12-S15, S31). FactorySimPy violations (T8) also 
increase with detailed specifications, indicating that richer structural constraints 
raise the likelihood of framework-level violations.

\textbf{Impact of description granularity.}
Figures~\ref{fig:error_type_frequency} and~\ref{fig:error_comp} compare error profiles 
between coarse and detailed descriptions for the same models. Providing detailed 
descriptions generally reduces total error counts, but the improvement is not uniform. 
While detailed descriptions effectively reduce naming errors (T1) by constraining 
component identifiers, they also introduce new error sources, particularly hierarchy 
mismatches (T6) and FactorySimPy violations (T8). This may reflect inherent limitations 
of natural language for describing networks and graphs: natural language is often 
context-dependent and ambiguous when specifying connectivity patterns, making detailed 
structural descriptions prone to misinterpretation. Increased detail shifts the error 
profile rather than eliminating errors entirely. Coarse descriptions produce errors 
dominated by naming (T1) and hallucinations (T3-T5), reflecting underspecification. 
Detailed descriptions exhibit more balanced error distribution with increased 
contributions from hierarchy (T6) and constraint violations (T8).

\textbf{Netlist versus Python as IR:}
To validate the density-preservation principle, we compared error rates between 
netlist-based IR (using Python dictionaries, as in our initial prototype) and 
Python-based IR for a subset of models. For S5, a simple linear sequence of 5 machines 
(7 nodes, 6 edges, 12 parameters), the netlist representation (13 dictionary entries) 
generated zero errors with detailed descriptions, indicating that enumerative IR 
remains adequate at small scale. However, as structural variation increased, netlist 
limitations became pronounced. For S8, a linear system with additional buffers inserted 
at specific positions, netlist-based IR produced 10 errors (structural hallucinations, 
incorrect parameter assignments, FactorySimPy constraint violations, improper edge 
chaining) while Python IR generated only 2 errors. For larger systems, the contrast 
is more dramatic. S34 (100 machines in series) expanded to over 205 netlist entries, 
and S35 (112 nodes, 165 edges with parallel and serial structures) required 278 entries. 
In S35, netlist-based IR hallucinated 10 nodes, 7 edges, and numerous unspecified 
parameter values due to explicit enumeration of repeated subsystems. Python IR, using 
loops and hierarchical structuring, encapsulated repetition compactly, substantially 
reducing hallucination errors. These results demonstrate that density-preserving IR 
reduces error accumulation, particularly for systems with regular or repeated structures. 
The compact loop-based Python IR also remains human-readable at these scales, whereas netlists with hundreds of entries become impractical to review by inspection.

\subsection{Key Observations}

\textbf{Naming errors and indexing ambiguity:}
Naming errors (T1) counterintuitively increase with description detail. The primary 
cause is indexing conflict: few-shot examples use 0-based indexing while natural 
language uses 1-based references. When descriptions refer to \texttt{Machine 1}, 
the LLM interprets this as index 1 rather than index 0, causing systematic off-by-one 
errors in grid systems (S21, S24). Buffers are frequently renamed to generic identifiers 
from examples (\texttt{edge[0]}, \texttt{src\_edge[0]}). Even explicit naming 
conventions like \texttt{Stage\_i\_M1} are inconsistently applied, especially with 
procedural rules (\texttt{for i from 1 to 3}) or spatial attributes (S6, S18, S30). 
Additional detail introduces more alignment opportunities rather than improving conformity.

\textbf{Parameter inference and defaults:}
When parameter specifications conflict, the LLM prioritizes specific over generic 
statements. In S13, stating "Machine in each stage has a processing delay of 4.0, 
3, and 2 seconds" overrides the later "All machines have a processing delay 
of 2". While internally consistent, this reflects implicit prioritization rather than 
explicit reasoning. When routing policies are unspecified, the LLM defaults to 
\texttt{ROUND\_ROBIN} (S30, S31-S35). In S23 and S25, routing instructions were 
hallucinated as explicit lists rather than mapped to supported FactorySimPy mechanisms 
(\texttt{ROUND\_ROBIN}, \texttt{FIRST\_AVAILABLE}, functions).

\textbf{Hallucinations driven by constraint satisfaction:}
The LLM hallucinates nodes, edges, and parameters attempting to satisfy perceived 
constraints. In S3, given "SRC is connected to M1 and M1 to Sink via 
Buffers with IDs B1, B2", the LLM inferred two buffers between M1 and Sink and 
hallucinated an extra machine to satisfy the one-to-one edge constraint. In S18, 
"The output of the last node is fed to first machine in node1" was 
interpreted as connections to both machines in the target subsystem, creating two 
feedback loops instead of one. The LLM prioritizes structural rule satisfaction over 
faithful interpretation.

\textbf{Hierarchy collapse and global assumptions:}
The LLM frequently flattens nested subsystems into global topologies (S13, S15, S18, 
S24, S31), particularly for non-regular interconnections. In S26, hierarchical subsystems 
were reinterpreted as a single serial flow. Global sharing assumptions also occur: 
in S13 and S15, parallel sequences were assumed to share sources and sinks despite 
no explicit statement. In S13, "Every sequence of SRC is followed by a 
buffer \texttt{ID = B\_src\_1}" led to inference that one buffer \texttt{B\_src\_1} connects 
the source to all sequences (violating one-to-one edge constraints). In S24 (10x10 
grid), a single global source/sink was assumed instead of row-wise hierarchy. Positional 
reasoning also fails: in S30, "between the 2nd and 4th machine" was 
interpreted as replacement rather than insertion, leaving components disconnected.

\textbf{Structural regularity as resilience:}
Regular structures exhibit fewer errors than irregular ones regardless of size. S34 
(100 machines in series) shows almost no errors, while medium-sized irregular systems 
(S21, S22) and S24 (10x10 grid with absent machines) produce more errors. Among large 
systems (S31-S35), regular repeated structures (S32-S35) preserve hierarchy with minor 
naming drift, while S31 exhibits collapse. Linear scaling is robust; large-scale 
repetition with implicit hierarchy, irregular cross-connections, or positional semantics 
increases misinterpretation. Density-preserving Python IR effectively handles regular 
structures but does not fully mitigate errors from ambiguous hierarchical descriptions 
or non-regular patterns.

\subsection{Insights}

The error characterization reveals several key insights for designing resilient 
LLM-assisted modeling workflows. First, density-preserving intermediate representations 
substantially reduce errors: the transition from netlist to Python IR demonstrates 
that compact representations using loops and classes reduce hallucination opportunities, 
particularly for regular structures (S34: 100 machines, almost no errors with Python IR 
vs. 205+ netlist entries). Second, explicit structure beats implicit conventions: the 
LLM struggles with inferred hierarchy, positional semantics (\texttt{between 
machines 2 and 4}), and context-dependent organization, producing global assumptions 
(shared sources/sinks) when structure is underspecified. Third, description detail 
has non-uniform effects, reducing naming errors (T1) but increasing hierarchy mismatches 
(T6) and framework violations (T8), reflecting natural language's inherent ambiguity 
for describing networks. Fourth, structural regularity, not size, determines error 
likelihood: linear sequences and regular grids scale robustly (S32-S35) while medium-sized 
irregular systems (S21, S22, S26) produce disproportionately more errors. Fifth, 
component-based constraints effectively limit error surface: no Python syntax errors 
(T7) occurred across all experiments, confining errors to naming, parameters, 
hallucinations, and framework constraints that automated validation can detect. Sixth, 
the 0-based vs. 1-based indexing conflict is systematic and requires explicit alignment 
through few-shot examples, system prompts, or post-processing. Finally, implicit defaults 
emerge predictably (\texttt{ROUND\_ROBIN} routing, generic parameters) when specifications 
are incomplete. These insights suggest that effective LLM-assisted automation requires 
co-design of component libraries, intermediate representations, description templates, 
and validation mechanisms. While density-preservation, component-based composition, 
and orthogonalization create a robust foundation, human oversight through visualization, 
validation, and iterative refinement remains essential for trustworthy automation.

\subsection{Takeaways: Towards Effective Model Descriptions}

The error characterization reveals a balance between coarse and detailed descriptions. Coarse descriptions leave naming and component identity underspecified, leading to naming errors (T1) and hallucinations (T3, T4). Detailed descriptions constrain naming but introduce more hierarchy mismatches (T6) and framework constraint violations (T8) as the LLM is pushed to interpret richer structural cues. This error profiling study suggests directions in which a modeler could work around the typical errors and write more effective descriptions. First, \textbf{specify topology explicitly}. Clear articulation of component types, routing logic, and hierarchy reduces structural hallucinations. Second, \textbf{use library-matched terminology}. Employing standard manufacturing terms (machine, buffer, source, sink) aligned with the FactorySimPy component library reduces translation-induced errors. Third, \textbf{delegate parameters to DataFITR} when system data is available. Omitting numerical values from the description and letting DataFITR infer them from sensor data avoids parameter conflicts. To address residual naming inconsistencies even when the above guidelines are followed, FactoryFlow plans to support a post-hoc name mapping mechanism in which the user supplies a CSV mapping generated names to canonical names, applied automatically during validation. This decouples structural correctness from naming conventions and lets the LLM focus on topology. Quantitative validation of these guidelines across model types is planned as future work.

\section{Conclusions and Future Work} \label{sec:conclusion}

This paper addressed the challenge of building trustworthy LLM-assisted automation 
for manufacturing Digital Twin workflows, where resilience to hallucination and 
systematic human oversight are essential for practical adoption.
We introduced three design principles within FactoryFlow to ground LLM-generated 
models structurally, syntactically, and parametrically while preserving interpretability 
and scalability. First, we orthogonalize structural modeling and parameter fitting, 
separating LLM-driven structural synthesis from data-driven parameter inference. 
Second, we restrict model representation to interconnections of validated, parameterized 
component classes (FactorySimPy) rather than monolithic simulation code. Third, we 
employ Python as a density-preserving intermediate representation, leveraging loops, 
hierarchy, and composition to express repetition compactly and limit hallucination 
surface area.

Empirical characterization revealed key insights. Error rates 
correlate with structural complexity, not size: linear scaling is robust while 
multidimensional repetition, cross-subsystem coupling, and implicit hierarchy increase 
misinterpretation. These findings confirm that representational expansion 
and structural ambiguity drive hallucination accumulation, 
justifying the design principles suggested here.

\textbf{Limitations:}
We describe limitations in our work: some inherent to the approach, others related 
to the scope of our characterization study, and some representing ongoing work in 
the publicly available open-source implementation. The design principles target systems 
with static or sporadically changing structures. For dynamically evolving topologies 
where structural changes occur frequently and unpredictably, process mining or fully 
data-driven approaches may be better suited. Component-based modeling introduces some rigidity: behaviors not 
expressible through existing library components require extension. In future, LLM-based 
generation of raw simulation code from scratch may become sufficiently robust to 
eliminate the need for such architectural constraints. Recent trends suggest this may 
happen sooner than expected. However, explainability and interpretability 
will remain critical at least as long as human operators remain relevant. 
Regarding ongoing implementation work, automated structural change 
detection module is under development. The open-source release of DataFITR currently 
runs on historical data. Integration with real-time streaming data 
sources and seamless GUI integration are in progress. Automated design space exploration 
and optimization capabilities are planned as extensions in FactoryFlow. 
Concerning the characterization study, all experiments used a single LLM
backend (Gemini 2.5 Pro).  Comparative analysis across multiple LLM models, averaged across
multiple execution runs, cost-performance tradeoffs with smaller or locally hosted models and an expanded benchmark set
with heterogeneity metrics would be ideal for a more thorough characterization,
and is planned as future work. The study applies specifically to manufacturing systems representable within FactorySimPy's
component scope; generalizability to other simulation domains remains to be studied.

\textbf{Future Directions:}
Aside from addressing the limitations listed above, future directions involve several 
promising extensions. A concrete extension is to embed automated validation directly into the FactoryFlow pipeline, running after each model generation step. This includes KPI validators that compare throughput, cycle time, and resource utilization between the generated model and ground-truth data, and structural test suites that check connectivity, adherence to FactorySimPy constraints, and completeness of required components.
Extending FactorySimPy to additional manufacturing paradigms 
(material flows, hybrid models) and developing component libraries for other domains 
(logistics, healthcare, service systems) would be ideal to test generalizability. 

\textbf{In conclusion}, while challenges remain in achieving fully trustworthy automation, the combination 
of density-preserving intermediate representations, component-based composition, and 
systematic human oversight demonstrates that LLM-assisted workflows can be both 
powerful and practical when grounded in deliberate architectural choices that prioritize 
interpretability alongside automation.

\bibliographystyle{plain}
\bibliography{references}

\clearpage

\appendix

\section{Appendix: Error Taxonomy and Examples}

\begin{figure}[H]
  \includegraphics[width=\textwidth]{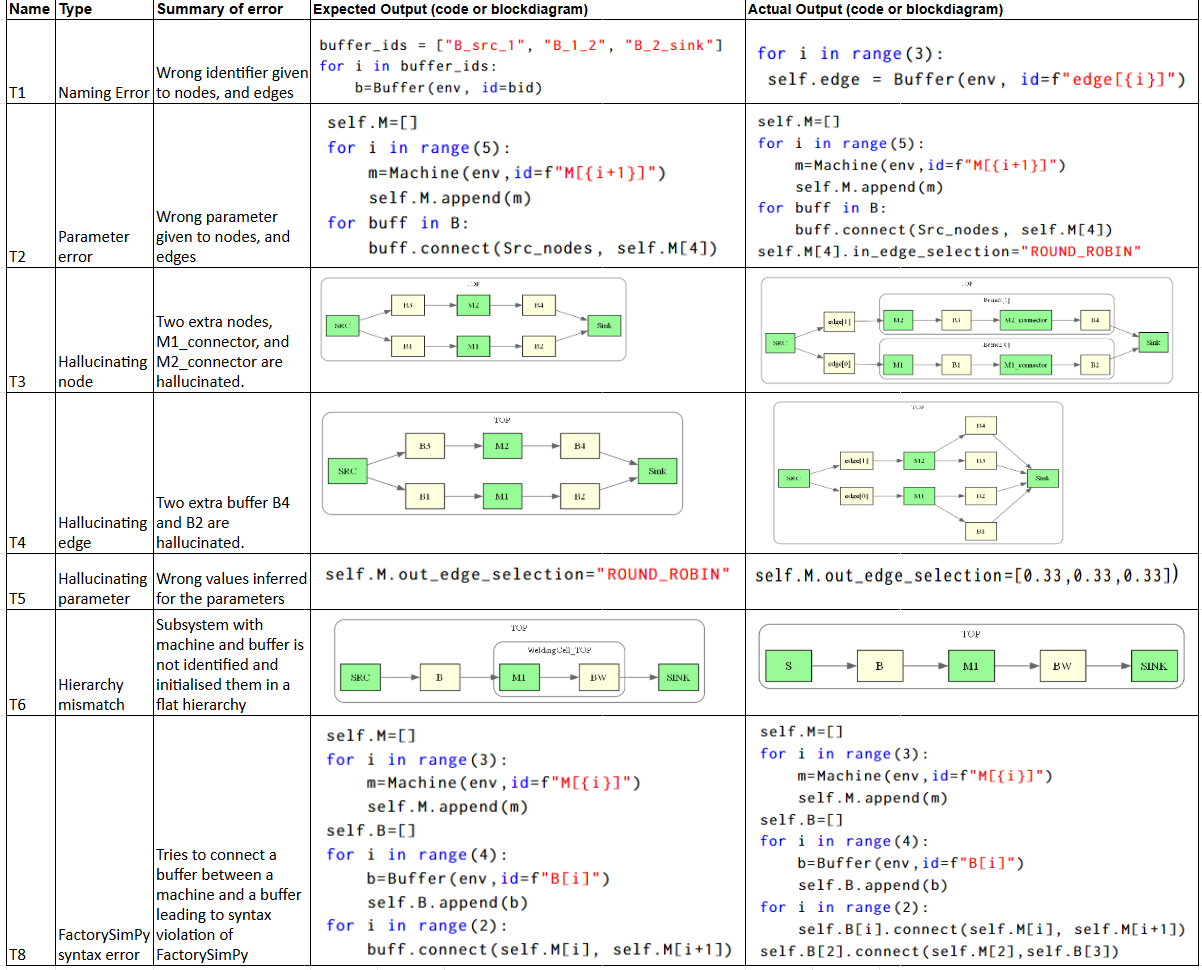}
  \caption{Examples of various types of errors observed}
\end{figure}

\clearpage
\section{Appendix: GUI Screenshots of the tools}

\begin{figure}[H]
\begin{center}
  \includegraphics[width=\textwidth]{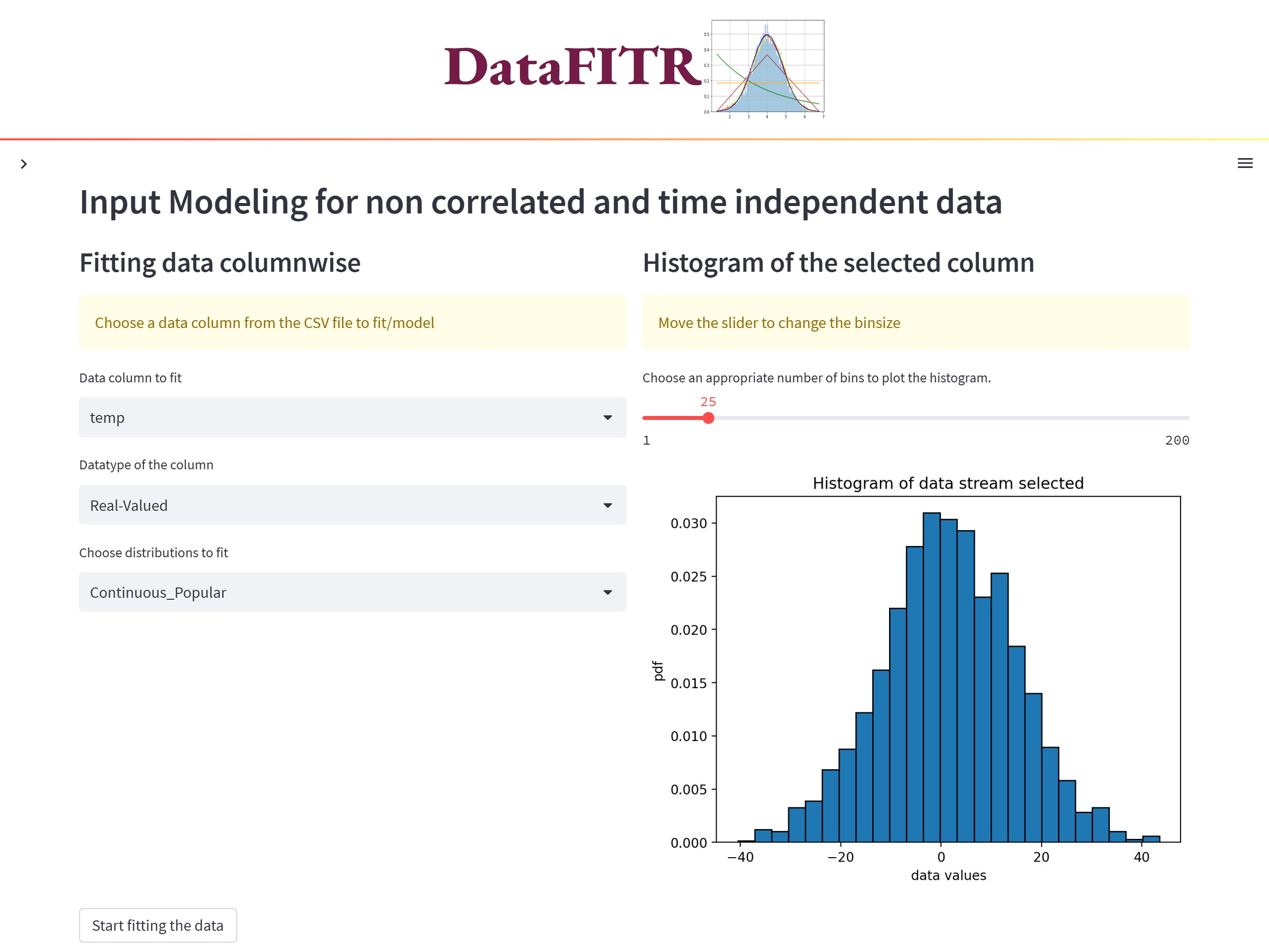}
  \caption{GUI of DataFITR (\href{https://datafitr.streamlit.app/}{DataFITR tool}) - Selecting data and fitting distributions.}
\end{center}
\end{figure}

\begin{figure}[H]
  \includegraphics[width=\textwidth]{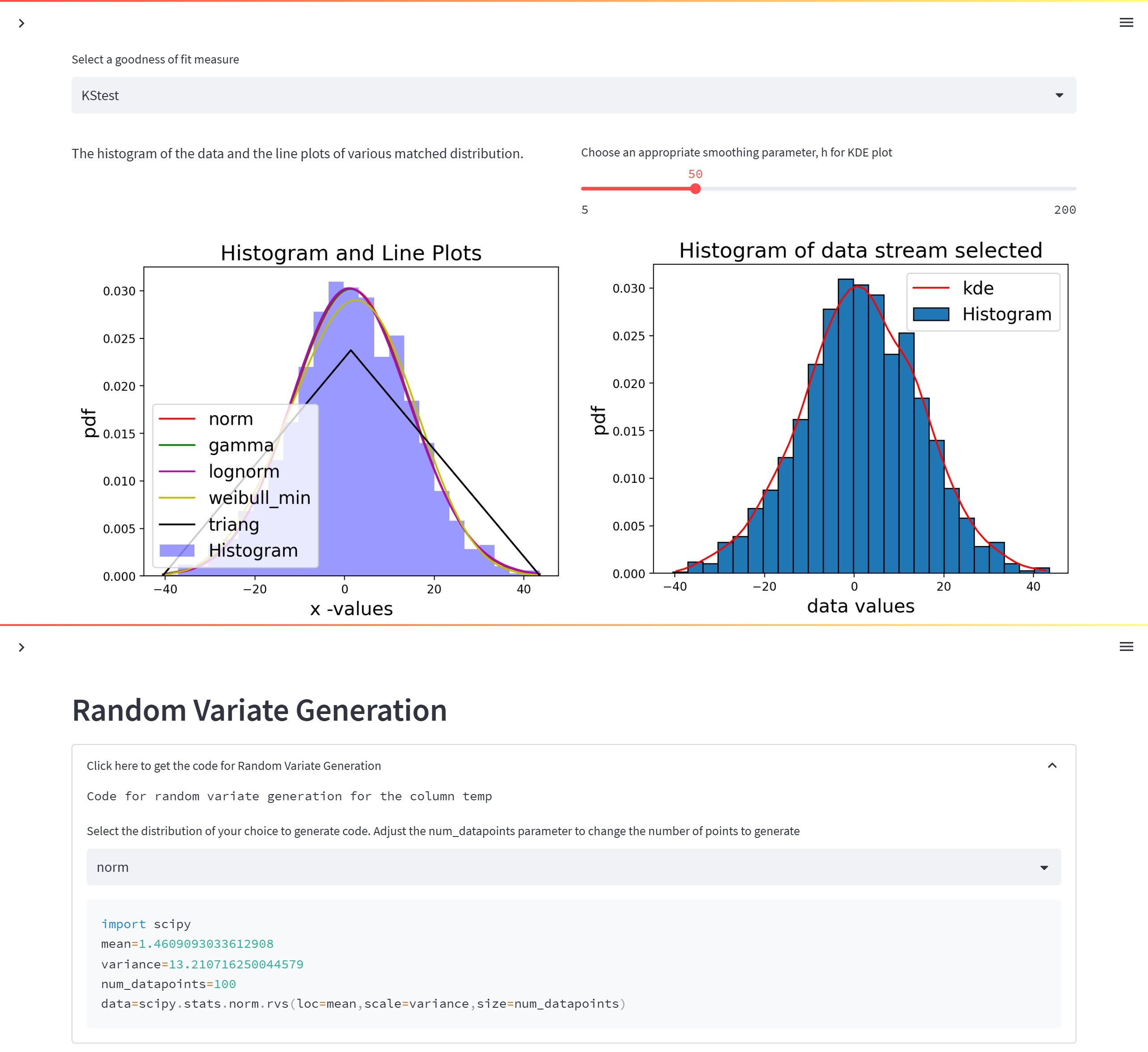}
  \caption{GUI of DataFITR - Results of input modeling.}
\end{figure}

\begin{figure}[H]
  \includegraphics[width=\textwidth]{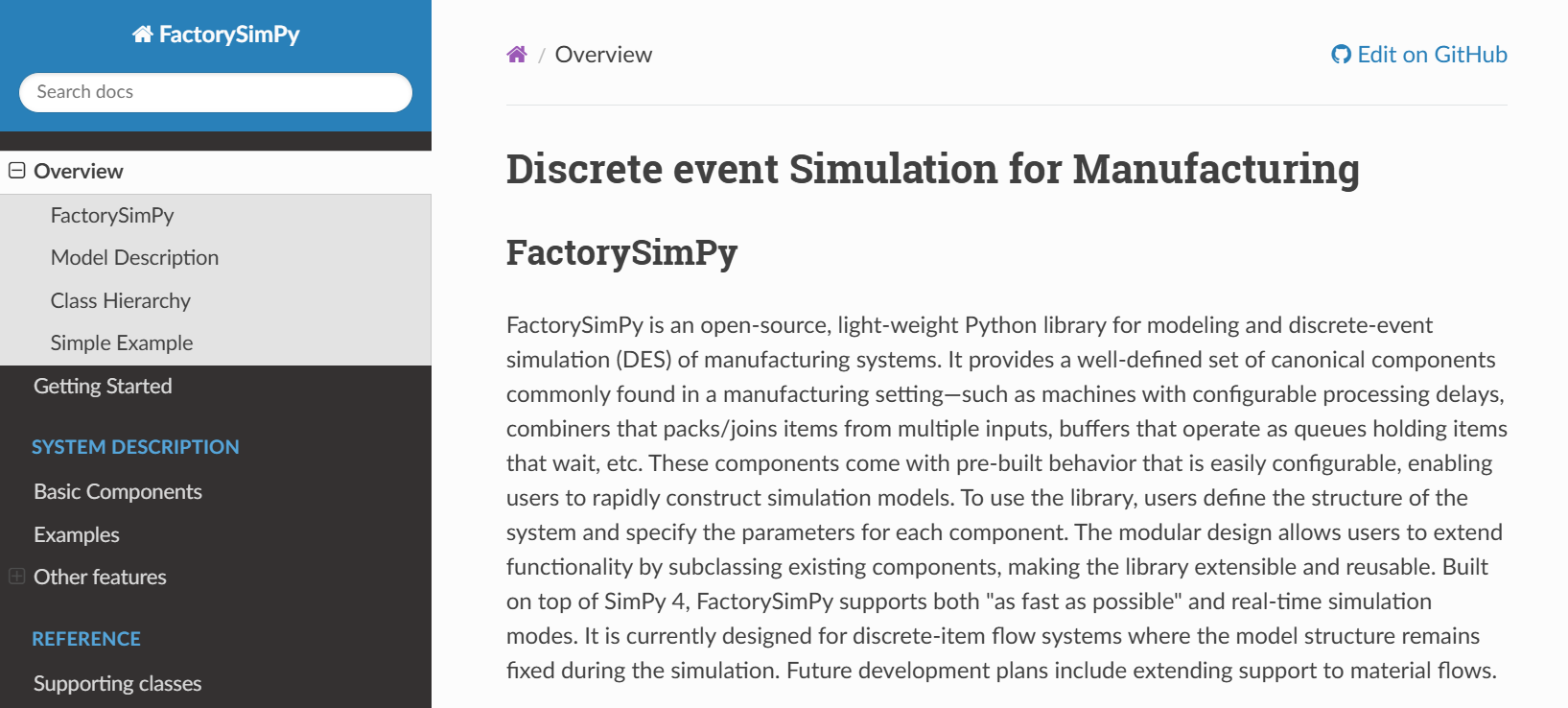}
  \caption{Documentation page of FactorySimPy (\href{https://factorysimpy.github.io/FactorySimPy}{documentation page}).}
\end{figure}

\begin{figure}[H]
  \includegraphics[width=\textwidth]{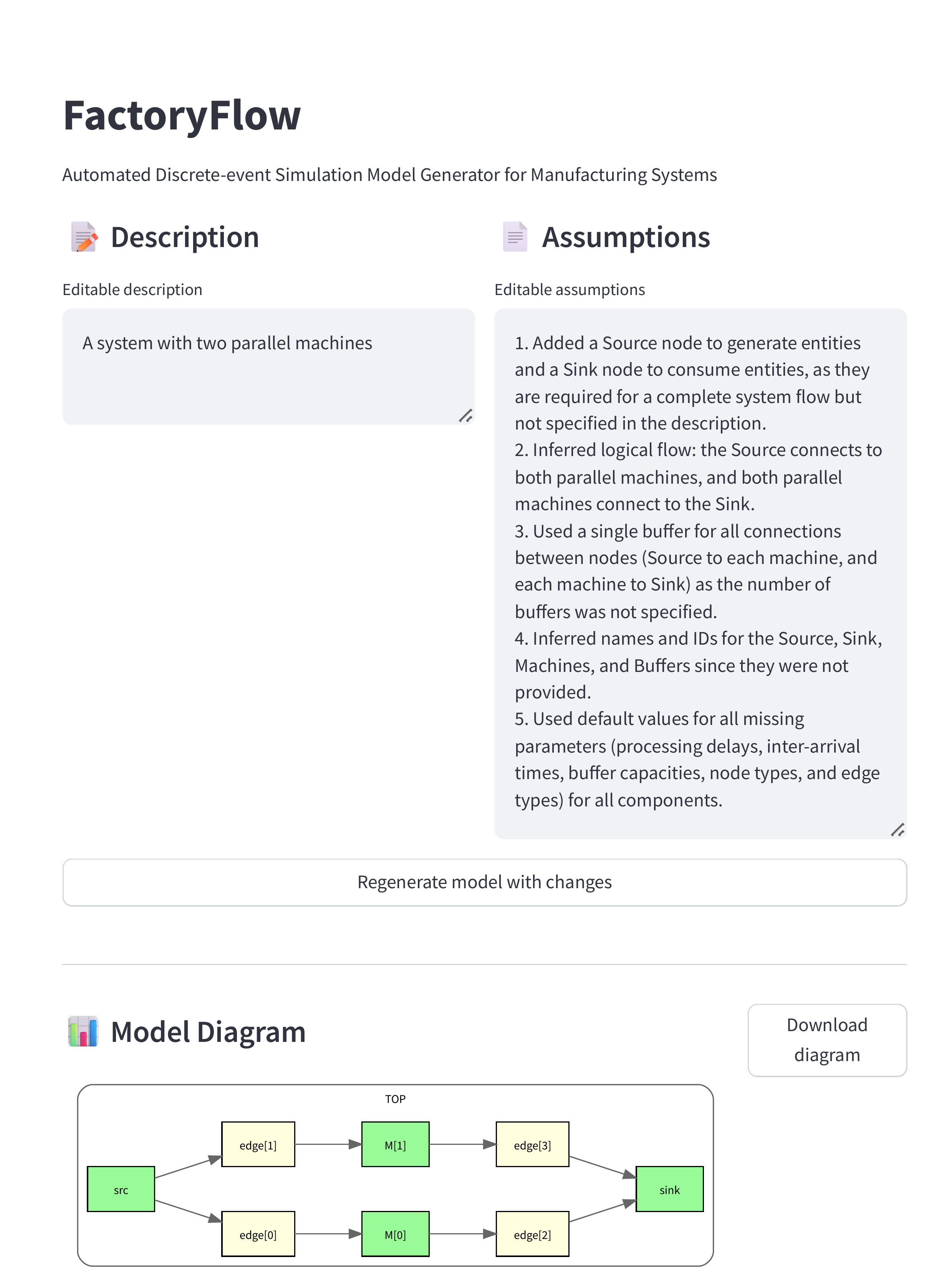}
  \caption{GUI of FactoryFlow (\href{https://github.com/InferaFactorySim/FactoryFlow}{GitHub repository (PoC)}), illustrating model generation of a simple system.}
\end{figure}

\begin{figure}[H]
  \includegraphics[width=\textwidth]{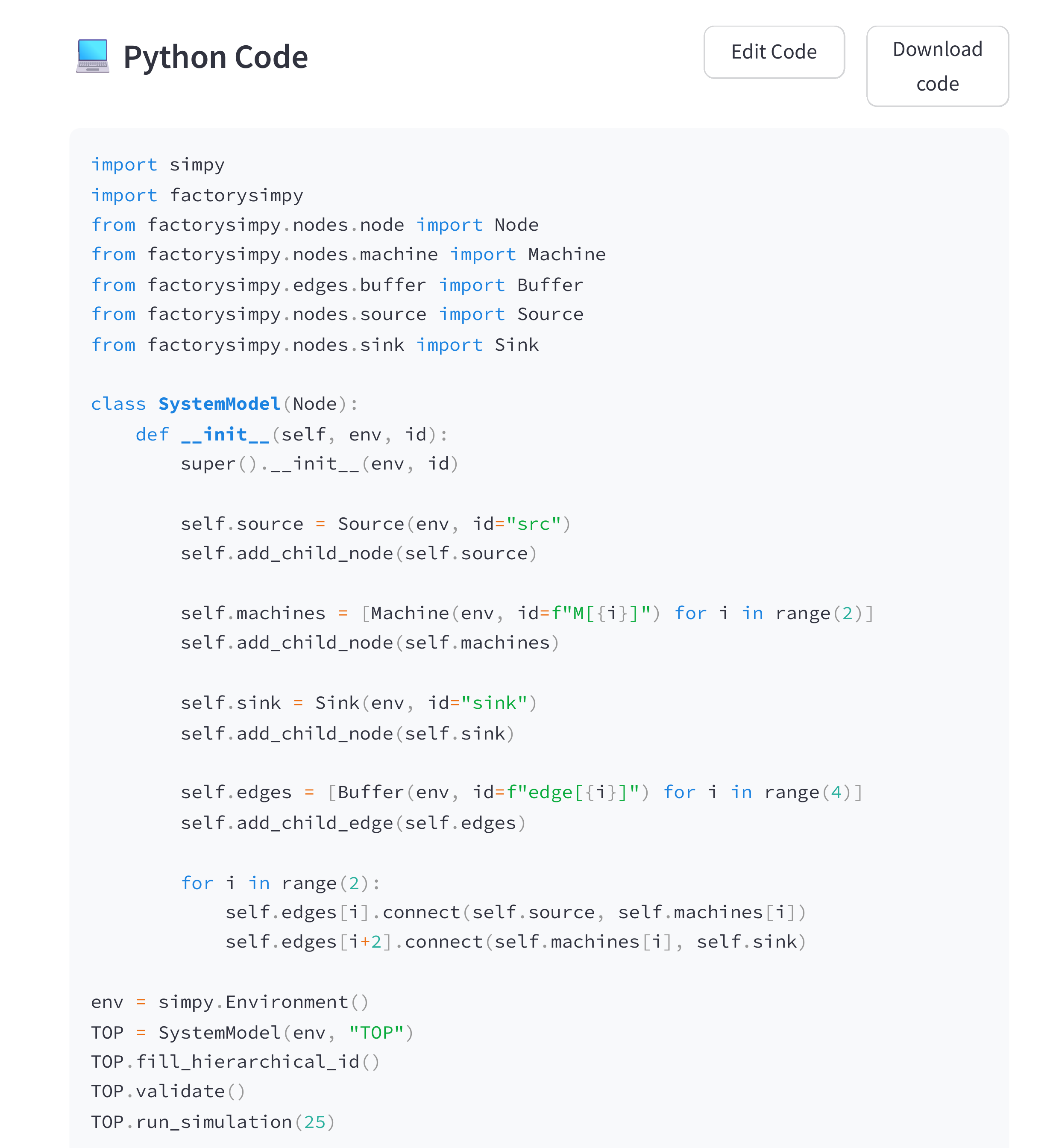}
  \caption{GUI of FactoryFlow with code generated for the description "A system with two machines in parallel"}
\end{figure}

\end{document}